\documentclass{article}
\usepackage{spconf,amsmath,graphicx,hyperref}
\usepackage{amssymb}
\usepackage{cite}
\usepackage{float}
\usepackage{xcolor}
\usepackage{multirow}
\usepackage{booktabs}
\usepackage{makecell}
\usepackage{subcaption} 
\usepackage{enumitem}
\usepackage{pgfplots}
\usepackage{makecell}
\usepackage{url}
\pgfplotsset{compat=1.18} 
\definecolor{darkgreen}{rgb}{0.0, 0.6, 0.0}
\definecolor{gold}{HTML}{FFCC00}

\title{Audio-Conditioned Diffusion LLMs for ASR and Deliberation Processing}
\name{Mengqi Wang$^{1\dag}$, Zhan Liu$^{2\dag}$, Zengrui Jin$^{2}$, 
      Guangzhi Sun$^{3}$, Chao Zhang$^{2}$, Philip C. Woodland$^{3}$
      \thanks{$\dag$ Equal contribution. Code and model are open-sourced at \url{https://github.com/liuzhan22/Diffusion-ASR}}}
      
\address{$^1$ University of Illinois at Urbana-Champaign,
         $^2$ Tsinghua University,
         $^3$ University of Cambridge \\
         \texttt{\small{mengqiw3@illinois.edu, liuzhan22@mails.tsinghua.edu.cn}}}
\begin{document}
\ninept
\maketitle
\begin{abstract}
Diffusion-based large language models (DLLMs) have recently attracted growing interest as an alternative to autoregressive decoders. In this work, we present an empirical study on using the diffusion-based large language model LLaDA for automatic speech recognition (ASR). We first investigate its use as an external deliberation-based processing module for Whisper-LLaMA transcripts. By leveraging the bidirectional attention and denoising capabilities of LLaDA, we explore random masking, lowest confidence masking, and semi-autoregressive strategies, showing that Whisper-LLaDA substantially reduces WER compared with the baseline. On LibriSpeech, the best cascade system achieves 2.25\%/4.94\% WER on test-clean/test-other, representing a 12.3\% relative improvement over the Whisper-LLaMA baseline on the test-other split. In contrast, a plain-text LLaDA without acoustic features fails to improve accuracy, highlighting the importance of audio-conditioned embeddings. We further evaluate Whisper-LLaDA as a standalone decoder for ASR with diffusion-based and semi-autoregressive decoding. Most experimental configurations achieve faster inference than the Whisper-LLaMA baseline, although recognition accuracy is slightly lower. These findings offer an empirical view of diffusion-based LLMs for ASR and point to promising directions for improvements. 
\end{abstract}
\section{Introduction}
\label{sec:intro}

Automatic speech recognition (ASR) has seen remarkable progress in recent decades~\cite{hadian2018end, gulati2020conformer,peng2022branchformer,yao2024zipformer}, driven largely by the growth in decoder model capacity~\cite{radford2023robust,tang2024salmonn,Qwen2-Audio,hurst2024gpt,Qwen2.5-Omni,abouelenin2025phi} and the adoption of monotonic and left-to-right autoregressive (AR) decoding, which sequentially generates tokens. However, despite their effectiveness, these approaches suffer from reduced efficiency and high computational cost due to the inherently sequential nature of AR decoding.

Several decoding strategies have been proposed to accelerate inference in ASR systems based on the attention encoder-decoder (AED) architecture, among which non-autoregressive (NAR) Transformer decoders have emerged as a widely adopted solution. By enabling greater parallelization compared with AR counterparts, NAR decoders significantly improve decoding efficiency and have attracted considerable attention in recent years~\cite{chen2020non,song2021non,yu2021non,zhang2022non}.

These NAR approaches can be broadly categorized as follows:
(1) Mask-based NAR systems \cite{higuchi2020mask,chan2020imputer,higuchi2021improved,zhang2023dynamic} are trained to iteratively reconstruct multiple masked tokens simultaneously, conditioned on those tokens that remain unmasked. 
(2) The hybrid AR-NAR strategy~\cite{tian2022hybrid,li2023improving,wang2024towards,arora2024semi,xu2025hainan} integrates both AR and NAR decoding mechanisms, thereby achieving a balance between the precision of AR models and the efficiency of NAR approaches. 
(3) Single-step connectionist temporal classification (CTC)~\cite{graves2006connectionist}-based approaches~\cite{song2021non,fan2021cass,deng2022improving} interpret entire token sequences in a single pass by leveraging decoder side CTC alignments and contextual information, resulting in a significant decoding speedup at the expense of a marginal degradation in recognition accuracy. 
(4) Alignment-level refinement methods have been developed, which either iteratively refine latent CTC alignments through multiple parallel denoising passes \cite{chi2020align}, or utilize ground-truth alignments to facilitate accurate sequence reconstruction in a single step \cite{chen2021align}.

However, despite substantially reducing decoding latency, existing NAR approaches generally suffer from a slight degradation in recognition accuracy compared to AR methods, reflecting an inherent trade-off between efficiency and performance. This motivates the search for alternative paradigms that have the potential to mitigate this trade-off. Recent advances in diffusion-based large language models (DLLMs)~\cite{austin2021structured,li2022diffusion,sahoo2024simple,shi2024simplified}, exemplified by LLaDA~\cite{nie2025large}, offer a promising direction. By combining powerful self-attention mechanisms with strong semantic modeling capabilities, DLLMs provide new opportunities for enhancing ASR. In this work, we present an empirical study on integrating DLLMs into ASR in two ways: (1) as external deliberation-based processing modules, where DLLMs refine preliminary hypotheses generated by a base recognizer; and (2) as internal decoders, directly replacing the conventional AR/NAR decoding backbone. These explorations highlight both the potential of DLLMs to reshape ASR decoding and the limitations that remain, pointing to important directions for future research.

\begin{figure}[t]
    \vspace{0.7em}
    \centering
    \includegraphics[width=\linewidth]{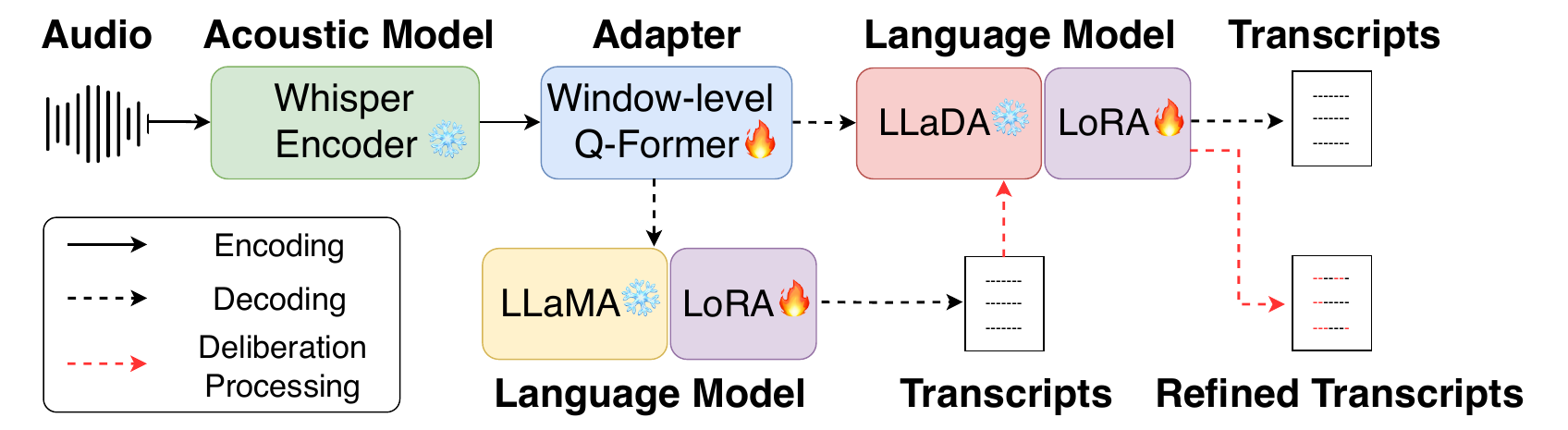}
    \caption{A flowchart illustrating the proposed ASR system with audio-conditioned LLaDA diffusion LLM used for deliberation processing.}
    \label{asr_sys}
    \vspace{-1.3em}
\end{figure}

Prior studies have explored external deliberation-based processing mechanisms that refine original hypotheses or N-best lists~\cite{wang2022deliberation,pandey2022lattention,li2023improving_icassp,kang2024transformer}, showing notable error reductions. Building upon these observations and inspired by recent progress in deliberation-based refinement, our study makes the following contributions: {\bf (i)} A systematic investigation on utilizing LLaDA as an external language model within ASR systems, demonstrating significant improvements in 
recognition accuracy compared to conventional autoregressive (AR) baselines.
{\bf (ii)} A novel exploration of employing LLaDA internally as a diffusion-based decoder within end-to-end ASR frameworks, highlighting substantial gains in parallel decoding capabilities and efficiency while maintaining competitive recognition performance relative to both AR and existing non-autoregressive (NAR) approaches.
{\bf (iii)} A systematic investigation of decoding strategies is conducted on the LibriSpeech corpus through extensive hyperparameter studies, specifically investigating a semi-autoregressive decoding strategy.

\vspace{-0.6em}
\section{Diffusion-based Large Language Models}
\vspace{-0.2em}

Diffusion-based large language models are defined by a generative process over discrete token sequences through a forward masking process and a learned reverse process. 
{\bf The forward process} gradually replaces tokens in a clean sequence $x_0$ with a mask token \texttt{[MASK]} as a function of continuous time $t\in[0,1]$, where the sequence is fully observed at $t=0$; each token is independently masked with probability $t$ as $t$ increases, and all tokens are masked at $t=1$.
{\bf The reverse process} starts from a fully masked sequence and iteratively predicts masked tokens conditioned on the unmasked ones, recovering a coherent sample at $t\to0$. 

LLaDA \cite{nie2025large} implements the forward and reverse process with a bidirectional attention-based Transformer, which predicts all masked positions in parallel. 
During the pre-training stage, LLaDA trains a mask predictor $p_{\theta}(\cdot \mid x_t)$ using a cross-entropy loss given by

{
\setlength{\abovedisplayskip}{1pt}
\setlength{\belowdisplayskip}{1pt}
\begin{equation}
    \mathcal{L}_{\operatorname{Pretrain}}(\theta)=-\mathbb{E}_{t, x_0, x_t}\left[\frac{1}{t} \sum_{i\in S_t} \log p_\theta\left(x_0^i \mid x_t\right)\right],
\end{equation}
}where $x_0 = [x_0^1, \cdots, x_0^i, \cdots,x_0^L]$ represents the clean sequence, $x_t$ is obtained by independently masking each position with probability $t \sim U(0,1]$, and $S_t$ is the set of positions of all masked tokens.
Supervised fine-tuning (SFT) is implemented using the following loss function:

{
\setlength{\abovedisplayskip}{1pt}
\setlength{\belowdisplayskip}{1pt}
\begin{equation}
    \mathcal{L}_{\operatorname{SFT}}(\theta)=-\mathbb{E}_{t, p_0, r_0, r_t}\left[\frac{1}{t} \sum_{i\in S_t^{\prime}} \log p_\theta\left(r_0^i \mid p_0, r_t\right)\right]
    \label{sft_loss},
\end{equation}
}where $(p_0, r_0)$ stands for paired prompt-response data, note that $p_0$ remains unmasked and $\mathcal{L}_{\operatorname{SFT}}$ is applied to the masked tokens of response $r_0$, whose positions are recorded in $S_t^{\prime}$. The formulation admits a time-free parameterization, thus the model can be trained to predict masked tokens given $x_t$ without taking $t$ as input, while still matching the diffusion semantics.

\vspace{-0.7em}
\section{Methods}
\vspace{-0.7em}
\label{sec:format}

\begin{figure*}
    \centering
    \includegraphics[width=\linewidth]{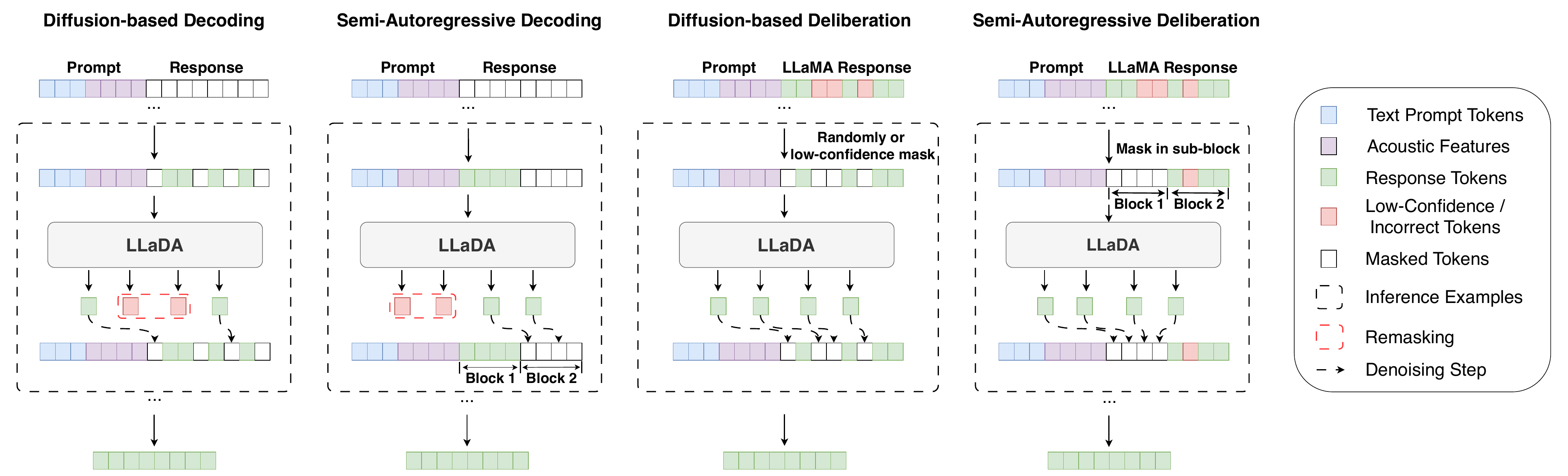}
    \caption{Overview of decoding and deliberation-based processing strategies.
(a) Diffusion-based decoding: generate the full response in parallel by iterative denoising.
(b) Semi-autoregressive decoding: split the response into sub-blocks, apply diffusion within each, and proceed autoregressively across sub-blocks.
(c) Diffusion-based deliberation: refine Whisper-LLaMA transcripts by randomly masking or masking lowest-confidence tokens, and then reconstructing them through diffusion.
(d) Semi-autoregressive deliberation: refine transcripts in sub-blocks, combining diffusion within each sub-block and autoregression across sub-blocks.}
    \label{inf}
    \vspace{-1.2em}
\end{figure*}

\subsection{Overall Architecture}

We propose \textbf{Whisper-LLaDA}, a model that integrates the Whisper-Large-v3 encoder~\cite{radford2023robust} with the diffusion-based large language model LLaDA-8B-Instruct~\cite{nie2025large}. The Whisper encoder extracts speech representations, which are passed through a window-level query Transformer (Q-Former)~\cite{li2023blip,tang2024salmonn}, which utilizes $4$ trainable queries with a $0.33$-second window, and a projection layer to align modalities.

The text decoder is LLaDA-8B-Instruct model, with LoRA~\cite{hu2022lora} applied to the query, key, and value projection layers of its self-attention blocks (rank $8$, scaling factor $4.0$, dropout $0.1$). Only LoRA, the Q-Former, and the projection layer are trainable with about $87$M parameters.

As shown in Figure~\ref{asr_sys}, Whisper-LLaDA supports both direct ASR from audio and deliberation-based processing of transcripts from external ASR systems, conditioned either on acoustic features alone or jointly on acoustic features and the given transcripts. We investigate this deliberation functionality on transcripts produced by a fine-tuned LLaMA-3.1-based~\cite{touvron2023llama} ASR system.

\vspace{-0.6em}
\subsection{Training}
We trained both Whisper-LLaMA and Whisper-LLaDA on the LibriSpeech corpus~\cite{panayotov2015librispeech}, which contains 960 hours of English audiobooks. Speed perturbation was performed with factors of 0.9 and 1.1. During training, the Whisper encoder is frozen, while LLaMA/LLaDA are fine-tuned with LoRA. The resulting speech embeddings are concatenated with the textual prompt and ground-truth response embeddings. Following the diffusion formulation, tokens in the response block are independently masked with probability $t \sim U(0,1]$, and LLaDA predicts the masked tokens conditioned on the remaining context. The model is optimized using cross-entropy loss over masked positions:
{
\setlength{\abovedisplayskip}{1pt}
\setlength{\belowdisplayskip}{1pt} 
\begin{equation}
    \mathcal{L}_(\theta)=-\mathbb{E}_{t, p_0, r_0, r_t,\alpha}\left[\frac{1}{t} \sum_{i\in S_t^{\prime}} \log p_\theta\left(r_0^i \mid p_0, r_t,\alpha\right)\right],
\end{equation}
}where $\alpha$ denotes the acoustic features of the input audio.

\vspace{-0.6em}
\subsection{Inference}
We organize the inputs into three components: a textual instruction, the acoustic features, and a response block. The textual instruction specifies the ASR task. The acoustic features are obtained by passing audio through the Whisper encoder, the Q-Former, and a projection layer, yielding 4096-dimensional embeddings aligned with LLaDA. For direct decoding, the response block is initialized with \texttt{[MASK]} tokens of length 128, which covers all LibriSpeech utterances as their lengths are shorter than 128 tokens. For deliberation-based processing, the response block is initialized with the Whisper-LLaMA transcript, and its length exactly matches that of the transcript sequence. LLaDA then performs the denoising process on this response block. Figure~\ref{inf} demonstrates the two decoding strategies considered in our work: diffusion-based and semi-autoregressive decoding, as well as two general deliberation-based processing strategies: diffusion-based and semi-autoregressive deliberation.

\vspace{-0.8em}
\subsubsection{Deliberation-based Processing with LLaDA}
Due to the denoising nature and bidirectional attention mechanism of the diffusion-based large language model, we explored a novel cascade framework for deliberation-based processing in ASR. We first obtain a transcript from the Whisper-LLaMA system to initialize the response block of Whisper-LLaDA. We select a proportion $p$ of the response tokens and replace them with \texttt{[MASK]} based on the following strategies: (1) random selection; (2) select the lowest confidence tokens, where confidence is obtained from a forward pass of Whisper-LLaDA based on the prompt and audio embeddings; and (3) semi-autoregressive deliberation-based processing: dividing the transcript into several sub-blocks and remasking sequentially. After remasking, Whisper-LLaDA recovers the masked positions conditioned on the input prompt, the speech embeddings, and the unmasked tokens of the transcript. The reconstruction under the first two strategies is completed in steps matching the mask count and the last entails sub-block-length steps per sub-block, closely aligning with our training procedures. For completeness, we also attempted a plain-text version of LLaDA without Whisper embeddings.

\vspace{-0.8em}
\subsubsection{Diffusion-based Decoding}
We define the total number of denoising steps as $N$, which is also the maximum number of iterations executed by LLaDA. In each iteration, LLaDA predicts all masked tokens in parallel, conditioned on the recovered ones, and produces confidence scores from the softmax distribution over logits. The top-$K$ tokens ($K = 128/N$) with the highest confidence are retained, while the remaining tokens are re-masked. This schedule guarantees the completion within $N$ steps.

To accelerate inference, we apply an early-stopping criterion: once an \texttt{[EOS]} token is denoised, all subsequent positions are forced to \texttt{[EOS]} to prevent redundant decoding. The decoding process ends once the entire response block is reconstructed. We evaluate with $N \in \{1, 4, 8, 16, 32, 64, 128\}$.

\vspace{-0.8em}

\subsubsection{Semi-autoregressive Decoding}
Beyond setting the maximum number of denoising steps $N$, the response block is divided equally into $M$ sub-blocks. Unlike the fully parallel strategy, the semi-autoregressive method performs diffusion-based decoding within each sub-block for up to $N/M$ steps, while proceeding autoregressively across sub-blocks. Specifically, we experiment with $M \in \{1,2,4,8,16\}$ and $N \in \{1,4,8,16,32,64,128\}$. The number of denoising steps per sub-block is constrained between 1 and $128/M$. The same early-stopping criterion is applied to improve efficiency.

\vspace{-0.3em}
\section{Experiments}
\vspace{-0.3em}
Our experiments were conducted on LibriSpeech \texttt{test-clean} and \texttt{test-other} sets~\cite{panayotov2015librispeech}. All models were trained with the AdamW~\cite{loshchilov2017decoupled} optimizer using a weight decay of 0.05. The learning rate followed a linear warmup–cosine decay schedule, starting from $1\times 10^{-6}$, increasing to $3\times 10^{-5}$ within 3000 steps, and decaying to a minimum of $1\times 10^{-5}$. The best checkpoint was selected based on the WER on the \texttt{dev-clean} set in each run.

\vspace{-0.5em}
\subsection{Baseline}
Table~\ref{tab:baseline} summarizes the baseline results on LibriSpeech benchmark. We primarily consider two LLM-based ASR systems, Whisper-LLaMA and Whisper-Vicuna~\cite{vicuna2023}, as our baselines. Both systems achieve strong recognition accuracy, with Whisper-LLaMA (Sys.~1) reaching 2.24\%/5.63\% WER on \texttt{test-clean}/\texttt{test-other} and Whisper-Vicuna (Sys.~2) showing slightly higher WER and a larger real-time factor (RTF).

We also report the performance of Whisper-Large-v2/v3 (Sys.~3–4). These systems achieve very competitive results (e.g., 3.90\% WER on test-other with Whisper-Large v3). We attribute this to their larger proprietary training datasets (up to $\sim$5M hours). Therefore, the results are provided only as reference values, as they are constructed in a manner that is not directly comparable to our systems.

\vspace{-0.5em}
\label{sec:pagestyle}
\begin{table}[htbp]
    \caption{Word error rate (WER\%) and real-time factor (RTF) for all proposed frameworks on LibriSpeech \texttt{test-clean/other}.}
    \footnotesize
    \setlength{\tabcolsep}{5pt}
    \centering
    \label{tab:baseline}
    \begin{tabular}{p{0.4cm}|c|cccc}
        \toprule
        \multirow{2}{*}{\makecell{Sys \\ No.}} & \multirow{2}{*}{\centering Model \& Setting} & \multicolumn{2}{c}{WER} & \multicolumn{2}{c}{RTF} \\
        \cmidrule(lr){3-4} \cmidrule(lr){5-6}
        & & clean & other & clean & other \\
        \midrule
        1 & Whisper-LLaMA 3.1 & 2.24 & 5.63 & 0.253 & 0.253 \\
        2 & Whisper-Vicuna    & 2.40 & 5.82 & 0.472 & 0.459 \\
        3 & Whisper-Large v2    & 2.87 & 5.16 & 0.196 & 0.216 \\
        4 & Whisper-Large v3    & 2.03 & 3.90 & 0.186 & 0.195 \\

        \midrule
          & \textbf{Whisper-LLaDA} &      &      &      &      \\
           & \hspace{1em}Step 1      & 11.04 & 17.56 & 0.033 & 0.039 \\
          & \hspace{1em}Step 4      & 5.37  & 10.72 & 0.046 & 0.052 \\
          & \hspace{1em}Step 8      & 3.78  & 7.39  & 0.055 & 0.063 \\
        5 & \hspace{1em}Step 16     & 3.13  & 6.32  & 0.073 & 0.080 \\
          & \hspace{1em}Step 32     & 2.96  & 5.80  & 0.112 & 0.122 \\
          & \hspace{1em}Step 64     & \textbf{2.82}  & 5.79  & 0.185 & 0.194 \\
          & \hspace{1em}Step 128    & 2.96  
        & \textbf{5.75}  & 0.333 & 0.343 \\

        \bottomrule
    \end{tabular}
    \vspace{-1.5em}
\end{table}

\subsection{Deliberation-based Processing}
We explored the use of LLaDA as an external deliberation processing module for transcripts generated by Whisper-LLaMA as follows.

First, we applied plain-text LLaDA to correct errors from Whisper-LLaMA. We initialized LLaDA with the LLaDA-8B-Instruct checkpoint and fine-tuned it on paired data consisting of Whisper-LLaMA transcripts and ground-truth text. The fine-tuned text-based LLaDA was then used to refine Whisper-LLaMA outputs. Notably, the response block length was set to match the length of the Whisper-LLaMA transcript. However, this approach introduced more errors, yielding 3.89\% WER on \texttt{test-clean} and 6.91\% on \texttt{test-other}, indicating that audio embeddings are essential for effective deliberation-based processing with LLaDA.

Subsequently, we turned to Whisper-LLaDA for deliberation-based processing on Whisper-LLaMA transcripts. In this setting, we directly used the end-to-end Whisper-LLaDA checkpoint without any additional training on Whisper-LLaMA outputs. Specifically, we investigated three strategies: (1) random masking; (2) lowest confidence tokens masking; and (3) semi-autoregressive paradigm. The results for strategies (1) and (2) are presented in Table~\ref{cascade_onestep}, while the results for strategy (3) are summarized in Table~\ref{cascade_chunks}.

\begin{table}[!t]
    \caption{WER (\%) of cascade deliberation-based processing with random/lowest confidence masking on LibriSpeech \texttt{dev}/\texttt{test} sets. ``Whisper-LLaMA" is the Whisper-LLaMA transcript baseline.}
    \label{cascade_onestep}
    \centering
    \footnotesize
    \setlength{\tabcolsep}{5pt}
    \begin{tabular}{c|cccc}
        \toprule
        \multirow{2}{*}{Mask ratio $p$} &
        \multicolumn{4}{c}{Random masking} \\
        \cmidrule(lr){2-5}
        & dev-clean & dev-other & test-clean & test-other \\
        \midrule
        Whisper-LLaMA & 2.24 & 5.07 & 2.24 & 5.63 \\
        10\%  & 2.25 & 5.05 & 2.23 & 5.59 \\
        30\%  & \textbf{2.21} & 5.02 & 2.23 & 5.59 \\
        50\%  & 2.28 & 5.05 & \textbf{2.21} & 5.55 \\
        70\%  & 2.31 & 4.98 & 2.22 & 5.37 \\
        90\%  & 2.32 & \textbf{4.92} & 2.23 & \textbf{5.24} \\
        100\% & 2.91 & 5.02 & 2.54 & 5.32 \\
        \midrule
        \multirow{2}{*}{Mask ratio $p$} &
        \multicolumn{4}{c}{Lowest confidence masking} \\
        \cmidrule(lr){2-5}
        & dev-clean & dev-other & test-clean & test-other \\
        \midrule
        Whisper-LLaMA & \textbf{2.24} & 5.07 & 2.24 & 5.63 \\
        10\%  & 2.27 & 5.06 & 2.24 & 5.63 \\
        30\%  & 2.28 & 5.03 & 2.21 & 5.52 \\
        50\%  & 2.32 & 5.02 & \textbf{2.18} & 5.43 \\
        70\%  & 2.34 & 4.97 & \textbf{2.18} & 5.34 \\
        90\%  & 2.31 & \textbf{4.96} & 2.26 & \textbf{5.23} \\
        100\% & 2.91 & 5.02 & 2.54 & 5.32 \\
        \bottomrule
    \end{tabular}
\end{table}

\begin{table}[!t]
    \caption{WER (\%) of cascade deliberation-based processing with different sub-block partitions on LibriSpeech \texttt{dev}/\texttt{test} sets. ``Whisper-LLaMA" is the Whisper-LLaMA transcript baseline.}
    \label{cascade_chunks}
    \centering
    \footnotesize
    \setlength{\tabcolsep}{5pt}
    \begin{tabular}{c|cccc}
        \toprule
        \multirow{2}{*}{\makecell[c]{Number of\\sub-blocks}} & \multicolumn{4}{c}{WER (\%)} \\
        \cmidrule(lr){2-5}
        & dev-clean & dev-other & test-clean & test-other \\
        \midrule
        Whisper-LLaMA & 2.24 & 5.07 & 2.24 & 5.63 \\
        2   & \textbf{2.21} & 4.83 & 2.25 & \textbf{4.94} \\
        4   & \textbf{2.21} & 4.84 & 2.22 & 5.21 \\
        6   & 2.26 & 4.86 & 2.23 & 5.26 \\
        8   & 2.30 & 4.86 & \textbf{2.20} & 5.25 \\
        10  & 2.27 & \textbf{4.81} & 2.21 & 5.25 \\
        \bottomrule
    \end{tabular}
    \vspace{-1em}
\end{table}

The deliberation-based processing results demonstrate that Whisper-LLaDA yields consistent improvements over the Whisper-LLaMA baseline, confirming the effectiveness of diffusion-based refinement in recovering tokens that are challenging for autoregressive counterparts. Random masking achieves the best results when the mask ratio is 90\%, while lowest confidence masking brings only marginal improvements. In addition, semi-autoregressive deliberation-based processing with sub-block partition further improves recognition accuracy, suggesting that the bidirectional attention of LLaDA provides a useful supplement to the unidirectional attention in the Whisper-LLaMA model. Overall, these findings highlight the potential of Whisper-LLaDA as a complementary deliberation-based processing module for ASR systems.

\vspace{-0.7em}
\subsection{Diffusion-based Decoding}
We observe the following trends for diffusion decoding (Sys.~5) in Table~\ref{tab:baseline}.
(a) Increasing the number of denoising steps steadily reduces WER, albeit with higher RTF. The best result on \texttt{test-clean} (2.82\%) is achieved with 64 steps, while the lowest WER on \texttt{test-other} (5.75\%) is obtained with 128 steps. These gains, however, diminish as the step count grows, indicating a saturation effect.
(b) The RTFs for 1–64 steps remain consistently lower than those of AR baselines. Notably, the 64-step setting achieves a favorable trade-off, with 2.82\%/5.79\% WER on \texttt{test\-clean}/\texttt{test-other}, surpassing Sys.~2 on the latter. It operates $1.3\times$ faster than Sys.~1 and $2.4\times$ faster than Sys.~2, offering more efficient inference with a moderate accuracy loss relative to most AR frameworks.

\begin{figure}[htbp]
    \centering
    \begin{subfigure}[b]{0.49\columnwidth}
        \centering
        \includegraphics[width=\linewidth]{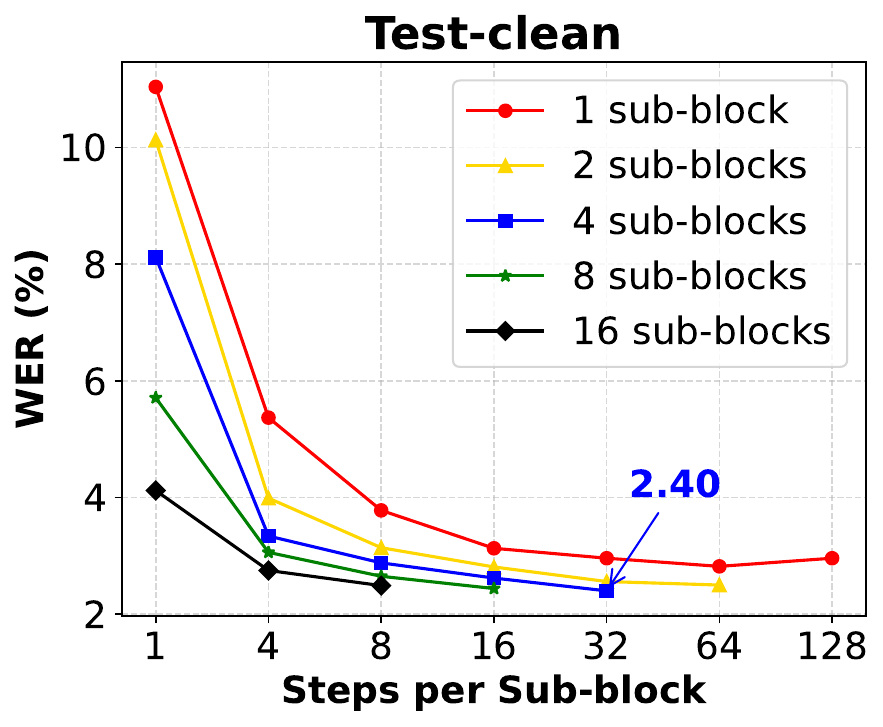}
        \caption{\texttt{test-clean}}
        \label{fig:clean}
    \end{subfigure}
    \hspace{-0.01\columnwidth}
    \begin{subfigure}[b]{0.49\columnwidth}
        \centering
        \includegraphics[width=\linewidth]{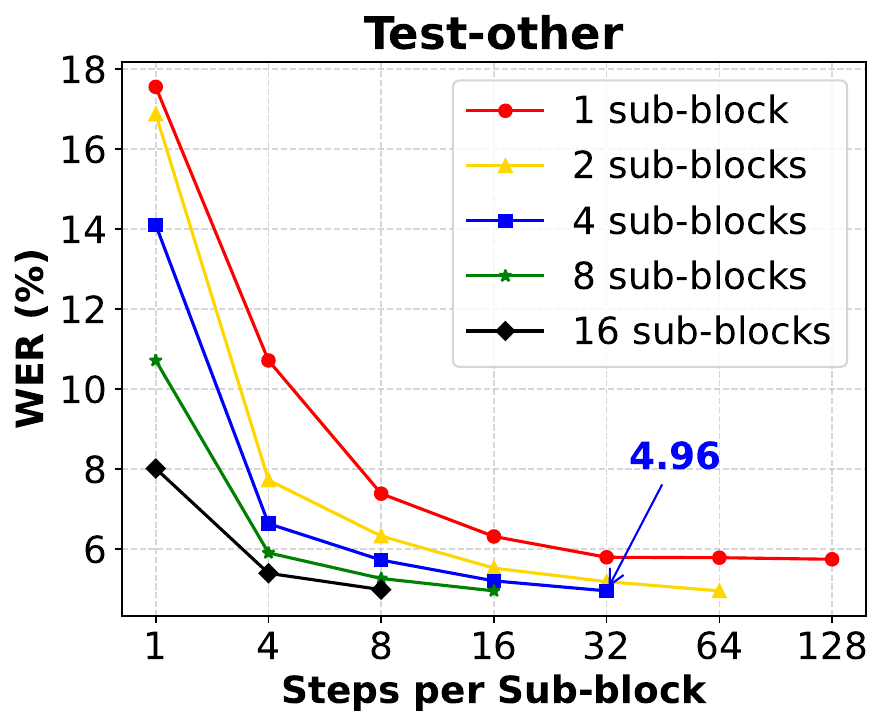}
        \caption{\texttt{test-other}}
        \label{fig:other}
    \end{subfigure}
    
    \caption{Effect of the number of denoising steps and the number of sub-blocks on WER for (a) \texttt{test-clean} and (b) \texttt{test-other}.}
    \label{semi}
    \vspace{-2em}
\end{figure}

\subsection{Semi-Autoregressive Decoding}
We investigate an alternative decoding strategy, namely semi-autoregressive decoding. The 128-token generation block is evenly divided into $M$ sub-blocks, within which LLaDA performs diffusion-based denoising, while proceeding in an autoregressive manner across sub-blocks. Up to 128 denoising steps are allocated and uniformly distributed among the sub-blocks. Figure~\ref{semi} shows the effect of the number of denoising steps and the number of sub-blocks on overall performance:
(a) Increasing the number of denoising steps per sub-block reduces WER, though the improvement saturates after roughly 16 steps per sub-block.
(b) As the total number of steps grows, the performance gaps among settings become negligible.
(c) The 4-block setting with 32 steps per sub-block achieves the best performance, with 2.40\%/4.96\% WER on \texttt{test-clean} and \texttt{test-other}. The performance on the latter is significantly better than the LLaMA/Vicuna-based ASR baselines.

\vspace{-0.1em}

\vspace{-0.4em}
\section{Conclusion}
\vspace{-0.4em}
This work presented a comprehensive empirical study of employing the diffusion-based language model LLaDA for automatic speech recognition. We examined three complementary perspectives: (i) applying plain-text LLaDA for transcript refinement, (ii) leveraging Whisper-LLaDA for cascade deliberation-based processing, and (iii) using LLaDA as a standalone decoder. Our findings reveal clear contrasts across these settings. Plain-text LLaDA fails to improve Whisper-LLaMA outputs, underscoring the necessity of audio embeddings for effective deliberation-based processing. In contrast, Whisper-LLaDA consistently enhances recognition quality, with random masking at high ratios and the semi-autoregressive framework delivering the largest gains. When used directly as a decoder, LLaDA enables faster inference than autoregressive baselines but with slightly higher WER in most settings.

Overall, these results highlight both the potential and the current limitations of diffusion-based models in ASR. LLaDA offers effective deliberation-based processing and efficient decoding when conditioned on acoustic features, yet its accuracy still lags behind extensively pretrained autoregressive systems. Future work should scale Whisper-LLaDA with larger and more diverse training data beyond LibriSpeech and systematically explore more advanced masking/remasking policies to narrow the gap while preserving efficiency.
\vfill\pagebreak

\begingroup
\footnotesize

\bibliographystyle{IEEEbib}
\bibliography{refs}

@inproceedings{yao2024zipformer,
  title={Zipformer: A faster and better encoder for automatic speech recognition},
  author={Z. Yao and L. Guo and X. Yang and W. Kang and F. Kuang and Y. Yang and Z. Jin and L. Lin and D. Povey},
  booktitle={Proc. ICLR},
  address={Vienna},
  year={2024}
}

@inproceedings{gulati2020conformer,
  title     = {Conformer: Convolution-augmented Transformer for Speech Recognition},
  author    = {A.~Gulati and J.~Qin and C.-C.~Chiu and N.~Parmar and others},
  year      = {2020},
  booktitle = {Proc. Interspeech},
  address   = {Shanghai}
}

@inproceedings{radford2023robust,
  title={Robust speech recognition via large-scale weak supervision},
  author={A.~Radford and J.~W.~Kim and T.~Xu and G.~Brockman and C.~McLeavey and I.~Sutskever},
  booktitle={Proc. ICML},
  address={Honolulu},
  year={2023}
}

@inproceedings{tang2024salmonn,
  title={{SALMONN}: Towards Generic Hearing Abilities for Large Language Models},
  author={C.~Tang and W.~Yu and G.~Sun and X.~Chen and T.~Tan and W.~Li and L.~Lu and Z.~Ma and C.~Zhang},
  booktitle={Proc. ICLR},
  address={Vienna},
  year={2024}
}

@article{Qwen2-Audio,
  title={Qwen2-Audio Technical Report},
  author={Y.~Chu and J.~Xu and Q.~Yang and H.~Wei and X.~Wei and Z.~Guo and Y.~Leng and Y.~Lv and J.~He and J.~Lin and C.~Zhou and J.~Zhou},
  journal={arXiv:2407.10759},
  year={2024}
}

@article{hurst2024gpt,
  title={{GPT}-4o system card},
  author={A.~Hurst and A.~Lerer and A.~P.~Goucher and A.~Perelman and A.~Ramesh and A.~Clark and A.~J.~Ostrow and A.~Welihinda and A.~Hayes and A.~Radford and others},
  journal={arXiv:2410.21276},
  year={2024}
}

@article{Qwen2.5-Omni,
  title   = {Qwen2.5-Omni Technical Report},
  author  = {J.~Xu and Z.~Guo and J.~He and H.~Hu and T.~He and S.~Bai and K.~Chen and J.~Wang and Y.~Fan and K.~Dang and B.~Zhang and X.~Wang and Y.~Chu and J.~Lin},
  journal = {arXiv:2503.20215},
  year    = {2025}
}

@article{abouelenin2025phi,
  title={{Phi-4-Mini} Technical Report: Compact yet Powerful Multimodal Language Models via Mixture-of-LoRAs},
  author={A.~Abouelenin and A.~Ashfaq and A.~Atkinson and H.~Awadalla and others},
  journal={arXiv:2503.01743},
  year={2025}
}

@inproceedings{graves2006connectionist,
  title={Connectionist Temporal Classification: Labelling Unsegmented Sequence Data with Recurrent Neural Networks},
  author={A.~Graves and S.~Fern{\'a}ndez and F.~Gomez and J.~Schmidhuber},
  booktitle={Proc. ICML},
  address={Pittsburgh},
  year={2006}
}

@inproceedings{higuchi2021improved,
  title={Improved Mask-{CTC} for Non-Autoregressive End-to-End {ASR}},
  author={Y.~Higuchi and H.~Inaguma and S.~Watanabe and T.~Ogawa and T.~Kobayashi},
  booktitle={Proc. ICASSP},
  year={2021},
  address={Toronto}
}

@inproceedings{higuchi2020mask,
  title={Mask {CTC}: Non-Autoregressive End-to-End {ASR} with {CTC} and Mask Predict},
  author={Y.~Higuchi and S.~Watanabe and N.~Chen and T.~Ogawa and T.~Kobayashi},
  booktitle={Proc. Interspeech},
  address={Shanghai},
  year={2020}
}

@inproceedings{wang2024towards,
  title={Towards Effective and Efficient Non-autoregressive Decoding Using Block-based Attention Mask},
  author={T.~Wang and X.~Xie and Z.~Li and others},
  booktitle={Proc. Interspeech},
  address={Kos},
  year={2024}
}

@article{tian2022hybrid,
  title={Hybrid autoregressive and non-autoregressive transformer models for speech recognition},
  author={Z.~Tian and J.~Yi and J.~Tao and S.~Zhang and Z.~Wen},
  journal={IEEE Signal Processing Letters},
  volume={29},
  pages={762--766},
  year={2022},
  publisher={IEEE}
}

@inproceedings{xu2025hainan,
  title={Three-in-One: Fast and Accurate Transducer for Hybrid-Autoregressive {ASR}},
  author={H.~Xu and T.~M.~Bartley and V.~Bataev and B.~Ginsburg},
  booktitle={Proc. ICLR},
  address={Singapore},
  year={2025}
}

@article{nie2025large,
  title={Large language diffusion models},
  author={S.~Nie and F.~Zhu and Z.~You and X.~Zhang and J.~Ou and J.~Hu and J.~Zhou and Y.~Lin and J.-R.~Wen and C.~Li},
  journal={arXiv:2502.09992},
  year={2025}
}

@inproceedings{hu2022lora,
  title={{LoRA}: Low-rank adaptation of large language models},
  author={E.~J.~Hu and Y.~Shen and P.~Wallis and Z.~Allen-Zhu and Y.~Li and S.~Wang and L.~Wang and W.~Chen and others},
  booktitle={Proc. ICLR},
  year={2022}
}

@inproceedings{fan2021cass,
  title={{CASS-NAT}: {CTC} alignment-based single step non-autoregressive transformer for speech recognition},
  author={R.~Fan and W.~Chu and P.~Chang and J.~Xiao},
  booktitle={Proc. ICASSP},
  address={Toronto},
  year={2021},
}

@inproceedings{song2021non,
  title={Non-autoregressive transformer {ASR} with {CTC}-enhanced decoder input},
  author={X.~Song and Z.~Wu and Y.~Huang and C.~Weng and D.~Su and H.~Meng},
  booktitle={Proc. ICASSP},
  address={Toronto},
  year={2021},
}

@article{chi2020align,
  title={Align-refine: Non-autoregressive speech recognition via iterative realignment},
  author={E.~A.~Chi and J.~Salazar and K.~Kirchhoff},
  journal={arXiv:2010.14233},
  year={2020}
}

@inproceedings{chen2021align,
  title={Align-Denoise: Single-Pass Non-Autoregressive Speech Recognition},
  author={N.~Chen and P.~Zelasko and L.~Moro-Vel{\'a}zquez and J.~Villalba and N.~Dehak},
  booktitle={Proc. Interspeech},
  address={Brno},
  year={2021}
}

@inproceedings{panayotov2015librispeech,
title={{LibriSpeech}: An {ASR} corpus based on public domain audio books},
  author={V.~Panayotov and G.~Chen and D.~Povey and S.~Khudanpur},
  booktitle={Proc. ICASSP},
  address={Brisbane},
  year={2015},
}

@inproceedings{chan2020imputer,
  title={Imputer: Sequence modelling via imputation and dynamic programming},
  author={W.~Chan and C.~Saharia and G.~Hinton and M.~Norouzi and N.~Jaitly},
  booktitle={Proc. ICML},
  address={Vienna},
  year={2020}
}

@inproceedings{zhang2023dynamic,
  title={Dynamic alignment mask {CTC}: Improved mask {CTC} with aligned cross entropy},
  author={X.~Zhang and H.~Tang and J.~Wang and N.~Cheng and J.~Luo and J.~Xiao},
  booktitle={Proc. ICASSP},
  address={Rhodes},
  year={2023},
}

@inproceedings{li2023improving,
  title={Improving non-autoregressive speech recognition with autoregressive pretraining},
  author={Y.~Li and L.~Samarakoon and I.~Fung},
  booktitle={Proc. ICASSP},
  address={Rhodes},
  year={2023},
}

@inproceedings{arora2024semi,
  title={Semi-autoregressive streaming {ASR} with label context},
  author={S.~Arora and G.~Saon and S.~Watanabe and B.~Kingsbury},
  booktitle={Proc. ICASSP},
  address={Seoul},
  year={2024}
}

@inproceedings{deng2022improving,
  title={Improving non-autoregressive end-to-end speech recognition with pre-trained acoustic and language models},
  author={K.~Deng and Z.~Yang and S.~Watanabe and Y.~Higuchi and G.~Cheng and P.~Zhang},
  booktitle={Proc. ICASSP},
  address={Singapore},
  year={2022}
}

@inproceedings{hadian2018end,
  title={End-to-end Speech Recognition Using Lattice-free {MMI}},
  author={H.~Hadian and H.~Sameti and D.~Povey and S.~Khudanpur},
  booktitle={Proc. Interspeech},
  address={Hyderabad},
  year={2018}
}

@inproceedings{peng2022branchformer,
  title={Branchformer: Parallel mlp-attention architectures to capture local and global context for speech recognition and understanding},
  author={Y.~Peng and S.~Dalmia and I.~Lane and S.~Watanabe},
  booktitle={Proc. ICML},
  address={Baltimore},
  year={2022}
}

@article{chen2020non,
  title={Non-autoregressive transformer for speech recognition},
  author={N.~Chen and S.~Watanabe and J.~Villalba and P.~{\.Z}elasko and N.~Dehak},
  journal={IEEE Signal Processing Letters},
  volume={28},
  pages={121--125},
  year={2020},
}

@inproceedings{zhang2022non,
  title={Non-autoregressive transformer with unified bidirectional decoder for automatic speech recognition},
  author={C.-F.~Zhang and Y.~Liu and T.-H.~Zhang and S.-L.~Chen and F.~Chen and X.-C.~Yin},
  booktitle={Proc. ICASSP},
  address={Singapore},
  year={2022},
}

@article{yu2021non,
  title={Non-autoregressive transformer-based end-to-end {ASR} using {BERT}},
  author={F.-H.~Yu and K.-Y.~Chen},
  journal={arXiv:2104.04805},
  year={2021}
}

@inproceedings{sahoo2024simple,
  title={Simple and effective masked diffusion language models},
  author={S.~Sahoo and M.~Arriola and Y.~Schiff and A.~Gokaslan and E.~Marroquin and J.~Chiu and A.~Rush and V.~Kuleshov},
  booktitle={Proc. NeurIPS},
  address={Vancouver},
  year={2024}
}

@inproceedings{li2022diffusion,
  title={{Diffusion-LM} improves controllable text generation},
  author={X.~Li and J.~Thickstun and I.~Gulrajani and P.~S.~Liang and T.~B.~Hashimoto},
  booktitle={Proc. NeurIPS},
  address={New Orleans},
  year={2022}
}

@inproceedings{austin2021structured,
  title={Structured denoising diffusion models in discrete state-spaces},
  author={J.~Austin and D.~D.~Johnson and J.~Ho and D.~Tarlow and R.~Van Den Berg},
  booktitle={Proc. NeurIPS},
  year={2021}
}

@article{loshchilov2017decoupled,
  title={Decoupled weight decay regularization},
  author={I.~Loshchilov and F.~Hutter},
  journal={arXiv:1711.05101},
  year={2017}
}

@inproceedings{li2023blip,
  title={{BLIP}-2: Bootstrapping language-image pre-training with frozen image encoders and large language models},
  author={J.~Li and D.~Li and S.~Savarese and S.~Hoi},
  booktitle={Proc. ICML},
  address={Honolulu},
  year={2023}
}

@inproceedings{li2023improving_icassp,
  title={Improving fast-slow encoder based transducer with streaming deliberation},
  author={K.~Li and J.~Mahadeokar and J.~Guo and Y.~Shi and G.~Keren and O.~Kalinli and M.~L.~Seltzer and D.~Le},
  booktitle={Proc. ICASSP},
  address={Rhodes},
  year={2023}
}

@inproceedings{pandey2022lattention,
  title={Lattention: Lattice-attention in {ASR} rescoring},
  author={P.~Pandey and S.~D.~Torres and A.~O.~Bayer and A.~Gandhe and V.~Leutnant},
  booktitle={Proc. ICASSP},
  address={Singapore},
  year={2022},
}

@inproceedings{kang2024transformer,
  title={Transformer-based model for {ASR} {N-best} rescoring and rewriting},
  author={I.~E.~Kang and C.~Van~Gysel and M.~H.~Siu},
  booktitle={Proc. Interspeech},
  address={Kos},
  year={2024}
}

@article{touvron2023llama,
  title   = {{LLaMA}: Open and efficient foundation language models},
  author  = {H.~Touvron and T.~Lavril and G.~Izacard and others},
  journal = {arXiv preprint arXiv:2302.13971},
  year    = {2023}
}

@misc{vicuna2023,
    title        = {Vicuna: An Open-Source Chatbot Impressing {GPT}-4 with 90\%* ChatGPT Quality},
    howpublished = {\url{https://lmsys.org/blog/2023-03-30-vicuna/}},
    author       = {W.-L.~Chiang and Z.~Li and Z.~Lin and Y.~Sheng and others},
    year         = {2023}
}

@inproceedings{wang2022deliberation,
  title={Deliberation of streaming rnn-transducer by non-autoregressive decoding},
  author={W. Wang and K. Hu and T. N. Sainath},
  booktitle={Proc. ICASSP},
  address={Singapore},
  year={2022},
}

@inproceedings{shi2024simplified,
  title={Simplified and generalized masked diffusion for discrete data},
  author={J.~Shi and K.~Han and Z.~Wang and A.~Doucet and M.~Titsias},
  booktitle={Proc. NeurIPS},
  address={Vancouver},
  year={2024}
}

\endgroup
\end{document}